\documentclass[%
 reprint, 
 amsmath,amssymb,
 aps
]{revtex4-2}

\usepackage{graphicx}% Include figure files
\usepackage{dcolumn}% Align table columns on decimal point
\usepackage{bm}% bold math
\usepackage{hyperref}% add hypertext capabilities
\begin{document}

\preprint{APS/123-QED}

\title{Large-scale mean flow and inclination of isolated turbulent band in channel flow}% Force line breaks with \\

\author{Linsen Zhang}
\author{Jianjun Tao}%
 \email{jjtao@pku.edu.cn}
\affiliation{%
 CAPT-HEDPS, SKLTCS,  Department of Mechanics and Engineering Science, College of Engineering, Peking University, Beijing 100871, China
 %This line break forced with %\textbackslash\textbackslash
}%


%\affiliation{
%\affiliation{
 %Third institution, the second for Charlie Author
%\affiliation{%


\date{\today}% It is always \today, today,

\begin{abstract}
Isolated turbulent bands observed in transitional channel flows have downstream heads and inclined bulks at a characteristic angle. In the large-scale mean flow, a $\nu$-shape vortex found at the head elongates into the bulk part, forming a pair of counter-rotating vortex tube structures. It is revealed numerically and theoretically that the head and the bulk convection velocities reflect the obliquely forward and the backward self-induced velocities of the $\nu$-shape vortex and the vortex tube pair, respectively, and the difference between these convection velocities provides a restoring angular momentum to retain the characteristic inclination angle through a self-adjustment process. 
\end{abstract}

\maketitle
Formation of localized turbulence in wall-bounded shear flows is one of the least understood phenomena in fluid dynamics \cite{tuckerman2020,avila2023,wu2023}.  As the precursor of the laminar-turbulent transition in plane-Poiseuille flow (pPF), isolated turbulent band (ITB) triggered by localized perturbations has been observed numerically \cite{tao2013, xiong2015, tao2018, kanazawa2018, shimizu2019,xiao2020} and experimentally \cite{liu2020, paranjape2020, mukund2021}, and illustrates fascinating features, e.g., its bulk part extends obliquely with a characteristic inclination angle. The oblique growth of turbulent patches is found to be related to the mismatch of streamwise flow rates  \cite{duguet2013}, and suitably tilted computational domains are used to simulate the transitional flows \cite{tuckerman2014, gome2020}. For plane-Couette flow (pCF), invariant solutions \cite{kawahara2012} bifurcated from the Nagata equilibrium \cite{Nagata90} are found to retain for a limited range of pattern angles \cite{reetz2019}, while the nonlinear traveling wave solutions obtained for pPF exist in a wide range of oblique angles \cite{paranjape2020}. When the Reynolds number decreases from a high value, alternating laminar and turbulent stripes or bands emerge from featureless turbulence in pCF \cite{Manneville11, manneville2012} and pPF \cite{kashyap2022}, but these periodic oblique bands are different from the ITB,  which has an energetic downstream end or ITB head and a decaying upstream tail, and is surrounded by a large area of laminar flow. Therefore, the problem: why the isolated turbulent band in pPF illustrates such a well-defined inclination angle, is thus still open.

\begin{figure}[h]
	\includegraphics[scale=0.05]{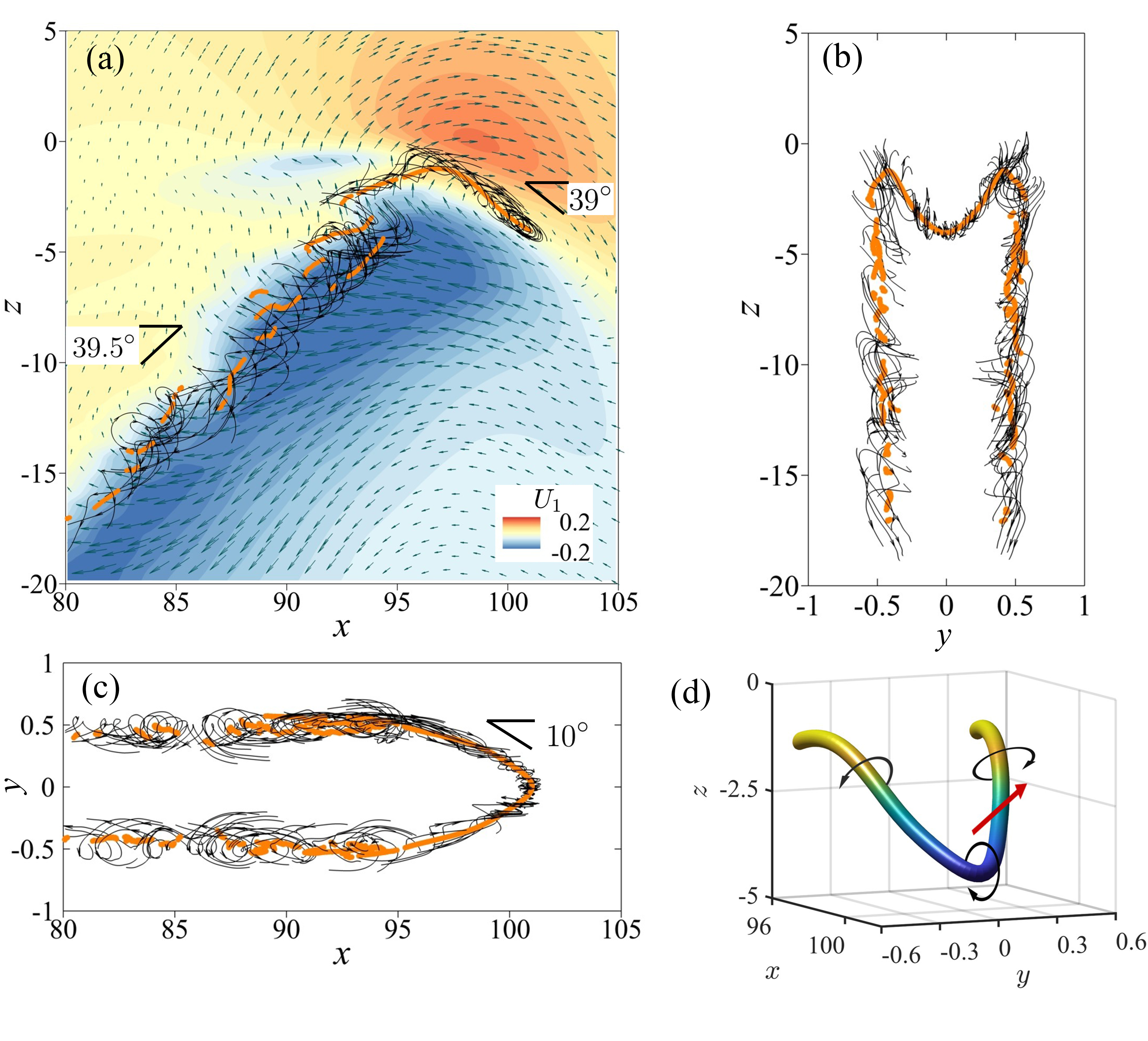}
	\caption{\label{fig1} Vortex cores (orange points)  in the mean-flow modulation ($\mathbf{U}_1$) field of ITB obtained at $Re =700$ in (a)  the bottom view, (b) the front view, and (c) the side view, where the streamlines (black curves) are truncated to avoid cloaking the cores. $U_1$ and $(U_1, W_1)$ fields at the midplane are shown by contours and vectors in (a), respectively.  (d) The vortex cores in the downstream head are linked into a $\nu$-shape vortex curve colored with $z$ coordinate, whose nearby induced velocity and self-induced velocity are illustrated with the black and red arrows, respectively.}
\end{figure}

The three-dimensional incompressible flow driven by a pressure gradient between two parallel walls is numerically simulated with a pseudo-spectral method \cite{simson}, where 1.5 times of the bulk velocity $U_m$ and the half channel height $h$ are set as the characteristic velocity and length scales, respectively. No-slip boundary conditions are applied at the walls ($y=\pm1$) and periodic boundary conditions are used in the streamwise ($x$) and spanwise  ($z$)  directions. The flow rate is kept constant and the flow field is expanded with Fourier modes ($N_x$ and $N_z$) and Chebyshev polynomials ($N_y$) in the $x$, $z$, and $y$ directions, respectively. The Reynolds number is defined as $Re=1.5U_m h/\nu$, where $\nu$ is the kinematic viscosity of fluid, and it is checked that the computational domain of $(\rm{L_x\times L_y\times L_z})=(200\times2\times160)$ is large enough and the resolution of $(\rm{N_x\times N_y\times N_z})=(1024\times65\times1024)$ with a time step of 0.02 is  fine enough to simulate the dynamic behaviors of ITB \cite{xiong2015,tao2018}.  

As shown in Fig.~\ref{fig1}(a), there is a high streamwise velocity region at the ITB head \cite{liu2020, kanazawa2018}, and hence the point with the maximum streamwise velocity at the midplane is named as head point and is tracked in the simulations to calculate its convection velocity  $\bold{u}_{\rm H}=(u_{\rm H},w_{\rm H})$, which  is referred as the head convection velocity hereafter. The velocity of ITB is decomposed into three parts within a frame moving with $\bold{u}_{\rm H}$, i.e., $\mathbf{U}=\mathbf{U}_0+\mathbf{U}_1+\mathbf{u}' $, where $\mathbf{U}_0$, $\mathbf{U}_1=(U_1,V_1, W_1)$, and $\mathbf{u}'$ are the velocities of the basic flow, the mean flow modulation, and the perturbations, respectively. At $Re=700$, the mean values of $(u_{\rm H},w_{\rm H})$ are calculated as $(0.8667,0.0994)$ with a convection angle  $\theta_{\rm H}={\rm arctan}(w_{\rm H}/u_{\rm H})=6.54^\circ$, and $\mathbf{U}_1$ is obtained by averaging 13,900 fields with a time interval of 0.02 as shown in Fig. 1, where the traveling streaks are filtered out.  Vortex cores, where $\nabla\mathbf{U}_1$ owns one real and a pair of complex-conjugate eigenvalues \cite{sujudi1995},  are calculated and linked into a bent smooth curve at the ITB head [Fig.~\ref{fig1}(d)], which is named as $\nu$-shape vortex hereafter. It extends into the bulk part, forming a pair of counter-rotating vortex-tube type structures or vortex tube pair near $y=\pm0.47$ as illustrated by the vortex cores in Fig.~\ref{fig1}(a)-\ref{fig1}(c). These large scale vortex structures make up the backbone of ITB. Note that in the frame co-moving with the ITB head,  the upstream tail extends obliquely with time and the mean flow near the tail is too weak to calculate the vortex core. 

\begin{figure}[h]
	\includegraphics[scale=0.059]{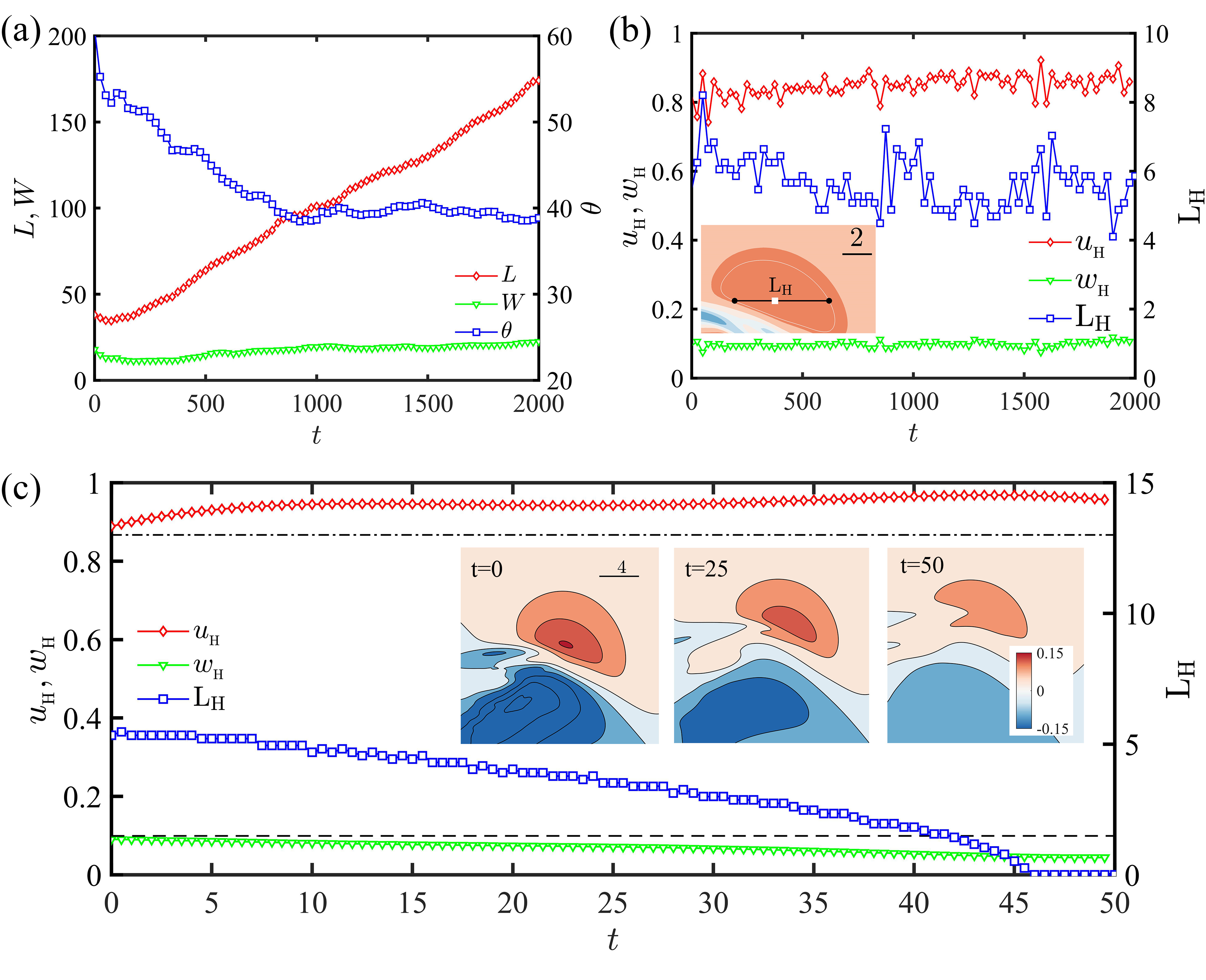}
	\caption{\label{fig2} (a) Geometric and (b) kinematic characteristics of the Oseen-type vortex initial disturbance measured every 1250 time steps during its evolution at $Re=700$. Inset of (b): the white point denotes the head point  and $L_{\rm H}$ is the length of the line segment where $U-U_0\ge 0.8(U-U_0)_{max}(t=0)$ at the midplane. (c) $\bold{u}_{\rm H}$ and $L_{\rm H}$ during the decay of the mean-flow structure shown in Fig. (1). As references,  $(u_{\rm H},w_{\rm H})$ of ITB are shown by the dash dot and dashed lines, respectively. Insets of (c): the iso-contours of $U-U_0$ around the downstream head at the midplane.}
\end{figure}

The  $\nu$-shape vortex has two important effects. The first is the induced flow at its neighborhood, e.g., the high speed region labeled by the red contours in Fig.~\ref{fig1}(a). The second is related to its curvature: similar to a vortex ring, the $\nu$-shape vortex should have an obliquely forward self-induced velocity as illustrated by the red arrow in Fig. 1(d). In order to verify this conjecture and inspired by the vortex structures shown in Fig.~\ref{fig1},  a wall-normal Oseen-type vortex superimposed with random noise is introduced  as the initial disturbance, 
\begin{equation}\label{eq:Oseen-initial}
	\begin{cases}
		u'_{\theta}&= f(y)\frac{\Gamma}{2\pi r}\left[ 1-{\rm exp}\left(-\frac{r^2}{R^2}\right)\right],\ v'=0\\ 
		f(y)&={\rm tanh}(\frac{1+y}{\kappa}) +{\rm tanh}(\frac{1-y}{\kappa})-1,
	\end{cases}
\end{equation} 
where $u'_\theta$ is the azimuthal velocity normal to the $y$-direction vortex axis and $r$ is the distance from the axis. The function $f(y)$ is used to damp the disturbance smoothly near the sidewalls, and the parameters ($\Gamma,R,\kappa)$ are set as $(8,2\sqrt{3},0.1)$ at $Re=700$ in order to mimic the mean flow around the $\nu$-shape vortex at the midplane.  This initial disturbance first evolves to a local structure similar as the ITB head at about  $t=150$, and then extends obliquely by producing quasi-periodic streaks to form the bulk part. The length $L$, width $W$, and inclination angle $\theta$ with respect to the $x$ direction of this evolving structure are calculated based on the centered second moments of disturbance kinetic energy relative to the basic flow, and for details of the calculation method we refer to the previous paper \cite{tao2018}. 

It is shown in Fig.~\ref{fig2}(a) that both $\theta$ and $W$ become statistically constant as $t>1000$, indicating that a mature ITB is developed. During the extension process, e.g., $t=(150, 2000)$ when $L$ increases from about 35 to 175 and $\theta$ decreases from about $53^\circ$ to $40^\circ$,  the segment length $L_{\rm H}$ shown in Fig.~\ref{fig2}(b), which represents the perturbation strength and length scale of the ITB head, fluctuates with time but around a constant value, and ($u_{\rm H}, w_{\rm H}$) remain statistically constant, suggesting that the head structure and convection velocities are independent of the characters of the bulk part, e.g., the band length and the inclination angle. Specifically, when traveling streaks are filtered out, it is shown in   Fig.~\ref{fig2}(c) that the large-scale mean flow continues to move downstream for some time, when both $u_{\rm H}$ and $w_{\rm H}$ still retain values close to those of the ITB, denoting that the head convection velocity is intrinsically determined by the mean flow, not by the perturbations, e.g., the traveling streaks. Note that without the Reynolds stress contributed by the perturbation components, the mean-flow modulation field decays due to the viscous diffusion and dissipation as illustrated in the insets of Fig. 2(c), and the high speed region at the downstream head shrinks as illustrated by the $L_{\rm H}$ curve.

\begin{figure}[h]
	\includegraphics[scale=0.062]{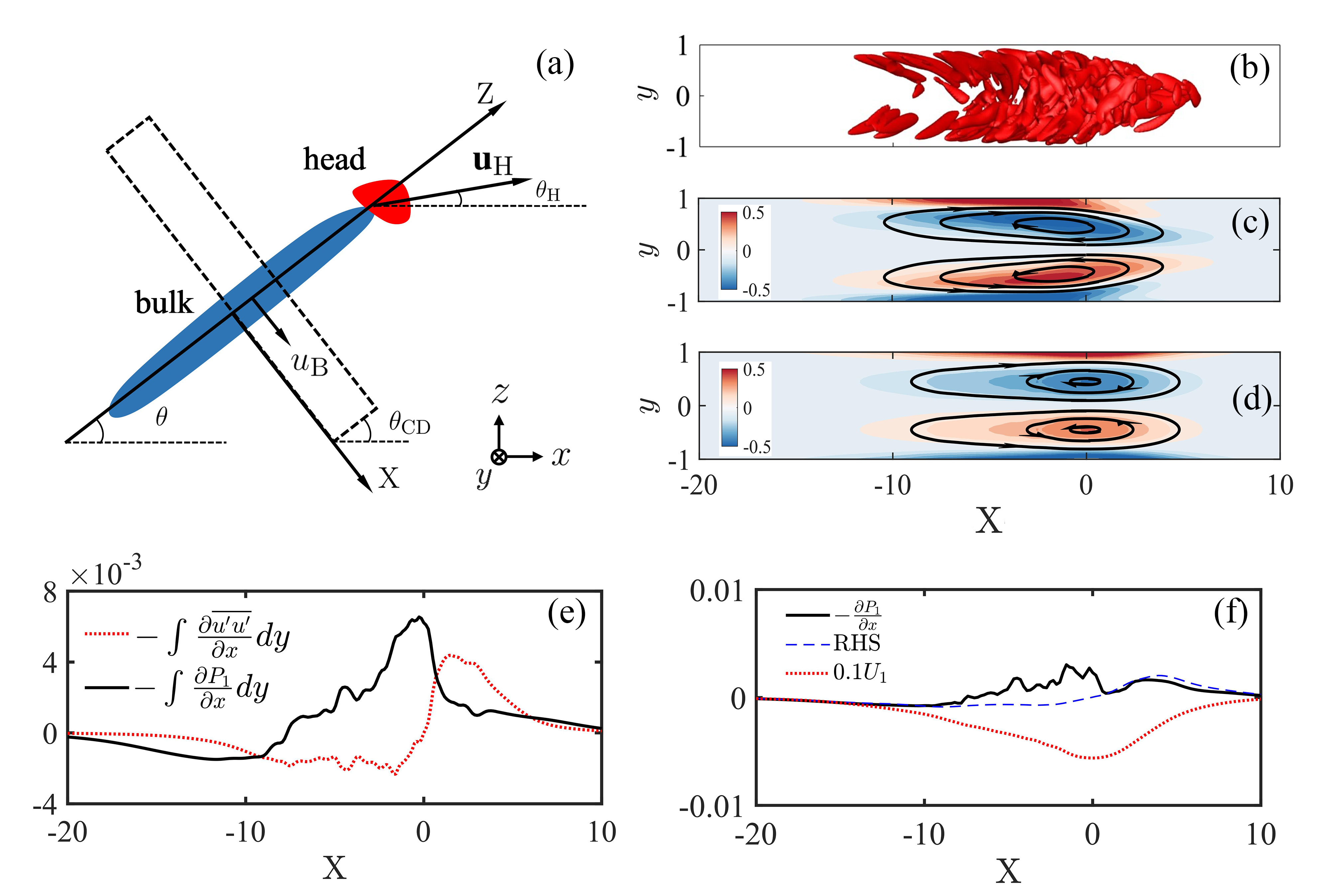}
	\caption{\label{fig3} Flow properties of BTB obtained at $Re=700$. (a) Bottom view of the tilted computational domain labeled with dashed lines in the tilted $X-Z$ coordinates. (b) Iso-surface of the vortex criterion $Q=0.02$ for transient perturbation field $\mathbf{u'}$. Iso-contours of $Z$-direction vorticity for the mean-flow modulation $\mathbf{U}_1$  obtained in simulations (c) and with the kinematic modal [Eq. (\ref{U1})] (d). The black streamlines illustrate the vortex tube pair. (e) Integrals of pressure gradient and Reynolds stress in Eq.  (\ref{int}). (f) Pressure term and the total right hand side (RHS) of Eq. (\ref{U_1}). In (c)-(f), the minimal $U_1$ at the midplane lies at $X=0$ . }
\end{figure}

A natural question is what determines the convection velocity of the ITB bulk. In order to answer this question, a long and tilted computational domain \cite{tuckerman2014} is used to simulate the bulk part as shown in Fig.~\ref{fig3}(a), which has no head and tail and is referred as the bulk turbulent band (BTB) hereafter. Note that by rotating the coordinates about the $x$-axis, an ITB with an inclination angle $\theta$ can convert to a counterpart with $-\theta$, and hence only positive tilted angles of the computational domain $\theta_{\rm CD}$  are considered, i.e., $0<\theta_{\rm CD}<90^{\circ}$. Periodic boundary conditions are applied in both the longitudinal ($Z$) and the transverse ($X$) directions of the band. The computational domain has the same width $L_{\rm Z}=10$ as the previous study \cite{paranjape2020} but a longer length ($L_{\rm X}=200$) to reduce the band interaction brought by the periodic boundary condition applied in the $X$ direction. Simulations are carried out with a resolution of $(\rm{N_X}\times\rm{N_y}\times\rm{N_Z})=(1024\times64\times64)$, which is checked to be fine enough to simulate the properties of BTB. 

The coordinate of BTB center $X_{\rm c}^{\rm B}$ is defined with the normal disturbance velocity $V(X,y,Z,t)$  at the midplane as $X_{\rm c}^{\rm B}=\int_0^{\rm L_X} \langle e_V\rangle_{\rm Z} X {\rm d}X /\int_0^{\rm L_X} \langle e_V\rangle_{\rm Z} {\rm d}X$, where $\langle e_V\rangle_{\rm Z}=\frac{1}{2L_{\rm Z}}\int_0^{L_{\rm Z}}V^2{\rm d}Z$, and then we can calculate the convection velocity of BTB, $u_{\rm B}={\rm d}X_{\rm c}^{\rm B}/{\rm d}t$.  The same velocity decomposition as that of ITB is applied  within a frame moving with  $u_{\rm B}$. It is noted that in the present configuration and parameter space, BTB can retain a saturated state with  a constant $u_{\rm B}$ and a steady $\mathbf{U}_1$ field for at least 3000 time units. At $Re=700$, $\mathbf{U}_1$ is obtained by averaging 20,000 fields with a time interval of 0.02, and the  vortex structures of  $\mathbf{u}'$ and  $\mathbf{U}_1$  are shown in Fig.~\ref{fig3}(b) and 3(c), respectively. The streamlines in Fig.~\ref{fig3}(c) illustrate clearly the vortex tube pair, which looks similar to the vortex dipole found in the mean flow of localized wave packet in two dimensional channel flows \cite{xiao2021} but with reverse rotating directions.  

Considering kinematic conditions for the mean flow: (a) no variation exists in the $Z$ direction and hence $\partial U_1 /\partial X+\partial V_1/\partial y=0$, (b) no-slip conditions at the walls, (c)  $\mathbf{U}_1$ is symmetric about the midplane, (d) the cross-section flow rate of  $\mathbf{U}_1$ is zero, and (e) $U_1(y=0)=U_1^c$, $(U_1, V_1)$  may be modeled with polynomials ($u_1, v_1$) as
\begin{equation}\label{U1}
%		\begin{cases}
	u_1=U_1^c(1-6y^2+5y^4), \ v_1=-\frac{dU_1^c}{dX}(y-2y^3+y^5).	
	%\end{cases}
\end{equation}
It is shown in Fig. 3(d) that this kinematic model grasps the main features of the vortex tube pair shown in Fig. 3(c), e.g., the vorticity directions.  The vortex centers predicted by the model lie at $y=\pm \sqrt{0.2}\simeq \pm 0.45$, which are close to $\pm 0.47$,  the simulation value for BTB. 

Next, the momentum equations governing the mean flow are simplified in order to estimate $u_{\rm B}$ analytically. As shown in Fig. 3(c), the length scale of  the vortex tube pair in the $X$ direction is much larger than that in the $y$ direction, and hence the pressure of the mean-flow modulation $P_1$ is  nearly independent of $y$, or approximately $\partial P_1/\partial X \simeq d P_1/dX$.  Integrating the time-averaged Navier-Stokes equation in the wall-normal direction and evaluating $(U_1, V_1)$ with $(u_1, v_1)$, we have
\begin{equation}\label{int}
	%		\begin{cases}
\int_{-1}^{1}\frac{\partial (P_1+\overline{u'u'})}{\partial X}dy\simeq -\frac{256}{315}\frac{d(U_1^c)^2}{d X}-\frac{64}{105}\frac{d U_1^c}{d X}+\frac{16}{Re} U_1^c,
		%\end{cases}
\end{equation}
 where the overline indicates the time average. Noticing that near the downstream end (e.g., $X>5$) $\int_{-1}^{1}\frac{\partial P_1}{\partial X}dy \simeq \int_{-1}^{1}\frac{\partial \overline{u'u'}}{\partial X}dy$ [Fig.~\ref{fig3}(e) ],  Eq.~(\ref{int}) is simplified as 
\begin{equation}\label{P}
	%		\begin{cases}
		\frac{\partial P_1}{\partial X}\simeq \frac{d P_1}{d X}\simeq -\frac{64}{315}\frac{\partial (U_1^c)^2}{\partial X}-\frac{16}{105}\frac{\partial U_1^c}{\partial X}+\frac{4}{Re} U_1^c.
		%\end{cases}
	\end{equation} 
At the midplane of the moving frame, the time-averaged $X$-direction momentum equation  is 
\begin{equation}\label{U_1}
	%		\begin{cases}
[sin(\theta_{\rm CD})-u_B+U_1^c]\frac{\partial U_1^c}{\partial X}=-\frac{\partial P_1}{\partial X}+\frac{1}{Re}\nabla^2 U_1+RS,
		%\end{cases}
	\end{equation}    
 where $RS$ represents the Reynolds stress term. It is shown  in Fig. \ref{fig3}(b) that the small-scale perturbation vortex structures concentrate near the midplane at the downstream side of BTB,  and then the corresponding Reynolds stress is localized in the $y$ direction as well. In order to retain the mean flow modulation near the midplane, $RS$ in Eq. (\ref{U_1}) should be balanced mainly by the viscous diffusion term instead of the pressure gradient, which is almost uniform in the $y$ direction. This conjecture is confirmed by the  numerical data shown in Fig. \ref{fig3}(f), where the right hand side (RHS) of Eq.~(\ref{U_1})  nearly coincides with $-\partial P_1/\partial X$ as $X>1$, indicating that $RS$ and the viscous diffusion term nearly cancel out each other near the downstream midplane. Consequently, $u_{\rm B}$ can be  evaluated analytically after substituting Eq.~(\ref{P}) into Eq.~(\ref{U_1}) as
  \begin{equation}\label{uB}
 		\begin{split}
u_{\rm B} & \simeq sin(\theta_{\rm CD})-\frac{16}{105}+\frac{187}{315}U_1^c + \frac{4U_1^c}{Re}  / \frac{\partial U_1^c}{\partial X}\\
&\simeq  sin(\theta_{\rm  CD})-\frac{16}{105}.
 		\end{split}
 \end{equation}       
Note that at the downstream midplane of BTB  as shown in Fig.~\ref{fig3}(f), $|U_1^c|\ll 1$, $ \frac{\partial U_1^c}{\partial X}>0$, and $| \frac{U_1^c}{Re}  / \frac{\partial U_1^c}{\partial X}|\ll 1$ due to $Re\gg 1$. At $Re=700$ and $\theta_{\rm  CD}=40^\circ $, Eq.~(\ref{uB}) predicts that $u_{\rm B}\simeq 0.49$,  which agrees well with the simulation value $0.48$, indicating that the present dynamic analyses grasp the key feature of the mean-flow structure. According to vorticity kinematics, the rotating directions of the vortex tube pair shown in Fig.~\ref{fig3}(c) correspond to a backward self-induced velocity, which may be estimated by deducting the basic flow velocity at the vortex center from $u_{\rm B}$ and the same value -0.02 is obtained for both the model and the simulations. 

\begin{figure}[h]
	\includegraphics[scale=0.031]{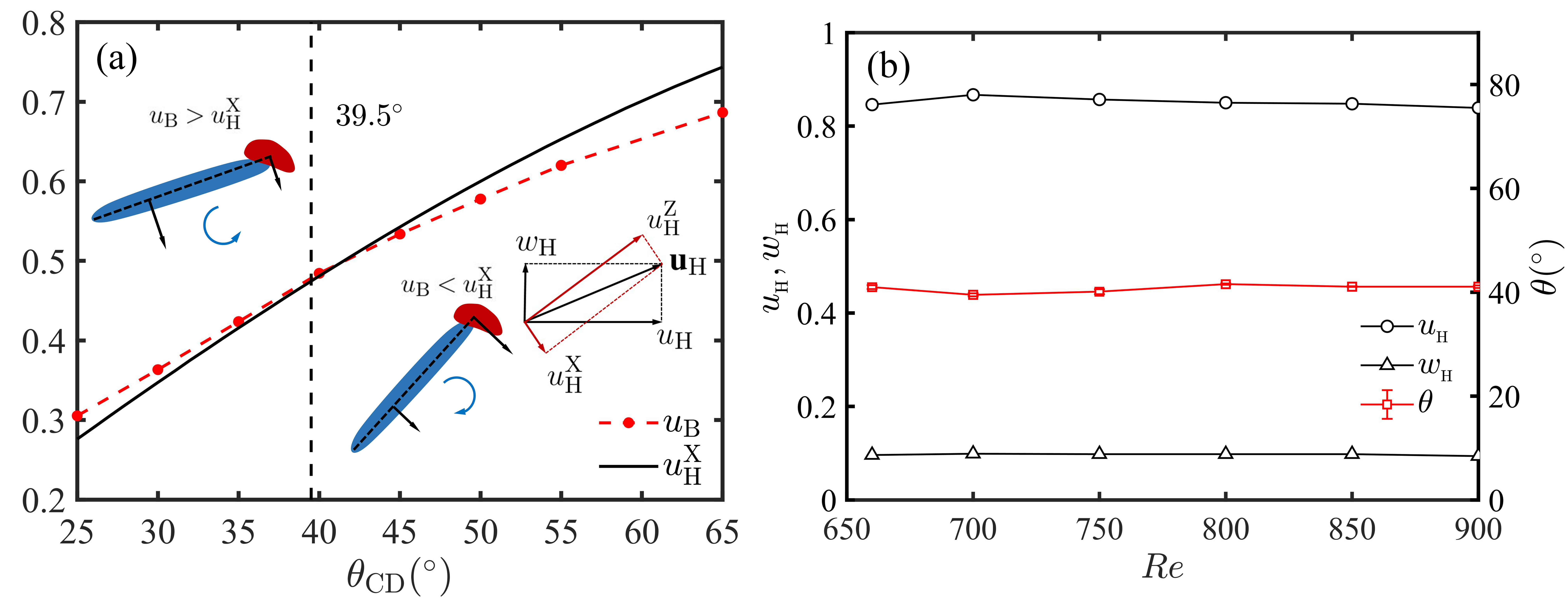}
	\caption{\label{fig4}(a) Convection velocity of BTB ($u_{\rm B}$) and the projection of the head convection velocity in the $X$ direction $u_{\rm H}^{\rm X}$ obtained numerically at $Re=700$ and different domain angles $\theta_{\rm CD}$. (b) Head convection velocity and inclination angle of ITB  as functions of $Re$.  }
\end{figure}

By comparing  $u_{\rm  B}$ obtained numerically at different $\theta_{\rm  CD}$ and $u_{\rm H}^{\rm X}=|\mathbf{u}_{\rm H}|{\rm sin}(\theta_{\rm CD}-\theta_{\rm H})$, the projection of ITB's head convection velocity  in the $X$ direction, an angle selection mechanism of ITB is proposed. It is shown in Fig.~\ref{fig4}(a) that when $\theta<39.5^\circ$, the corresponding $u_{\rm  B}$  is larger than $u_{\rm H}^{\rm X}$, and this velocity difference exerts an angular momentum to rotate anticlockwise the ITB during its movement, increasing the inclination angle $\theta$; while for $\theta>40.5^\circ$, $u_{\rm  B}<u_{\rm H}^{\rm X}$ and hence the ITB will rotate clockwise, leading to the decrease of  $\theta$. Therefore, the selection of ITB's inclination angle $\theta\simeq 40^\circ$ is a result of competition between  $u_{\rm  B}$ and  $u_{\rm H}^{\rm X}$ through a self-adjustment process. In fact, this process has been illustrated in Fig.~\ref{fig2}(a) , where the inclination angle of the vortex-triggered band structure decreases with time until it reaches the equilibrium value around $t=1000$. 

According to Eq.~(\ref{uB}), the bulk convection velocity is mainly determined by the large-scale mean flow and  is a weak function of $Re$, so is the head convection velocity shown in Fig.~\ref{fig4}(b). Consequently, the inclination angle of ITB is basically independent of $Re$ as well due to the self-adjustment mechanism and retains around $40^\circ$ [Fig.~\ref{fig4}(b)], a value consistent with the experimental measurements \cite{liu2020}. When turbulent bands interact with each other, e.g., the parallel arranged turbulent bands observed at high Reynolds numbers \cite{Tsukahara14}, their inclination angles may be different from that of the ITB. 

At low and moderate Reynolds numbers, the isolated turbulent band is the typical transitional structure of channel flows, and its most curious feature is probably the characteristic inclination angle of the bulk. Based on numerical simulations, kinematic modeling, and dynamic analyses, it is shown that the inclination angle is not determined by the bulk part itself, but by a balance between the convection velocities of the bulk and the downstream head: the difference between these convection velocities exerts an angular momentum to rotate the ITB until the specific inclination angle is achieved, indicating the crucial role played by the head part in the angle selection mechanism. This self-adjustment mechanism provides new insights into how localized turbulence behaves during the subcritical transitions in wall-bounded shear flows.

 The simulations are performed on TianHe-1(A), and the help on  SIMSON from P. Schlatter, L. Brandt, and D. Henningson is gratefully acknowledged. This work is supported by the National Natural Science Foundation of China (Grants Nos. 91752203).
%\nocite{*}
%\bibliographystyle{apsrev4-1} 
\bibliography{band}% Produces the bibliography via BibTeX.

\end{document}